\documentclass[letterpaper, 10 pt, conference]{ieeeconf}  

\IEEEoverridecommandlockouts                              
\usepackage{graphicx}      
\usepackage{amsmath}
\usepackage{amssymb}
\usepackage{mathrsfs}  
\usepackage{algorithm} 
\usepackage{hyperref}
\usepackage{enumerate}
\usepackage{xcolor}

\usepackage{lipsum}  

\usepackage{arydshln}
\usepackage{algpseudocode} 
\newcommand{\ip}[2]{\ensuremath\left\langle{#1}, {#2}\right\rangle}
\newcommand{\dom}{\ensuremath\mathrm{dom}\,}

\title{\LARGE \bf
Variable Metric Splitting Methods for Neuromorphic circuits simulation
}
\author{Amir Shahhosseini, Thomas Burger and Rodolphe Sepulchre
\thanks{A. Shahhosseini and Thomas Burger are with KU Leuven, Department of Electrical Engineering (STADIUS), KasteelPark Arenberg, 10, B-3001 Leuven, Belgium, {\tt\small amir.shahhosseini@kuleuven.be, thomas.burger@kuleuven.be}. R. Sepulchre is with both the KU Leuven, Department of Electrical Engineering (STADIUS), KasteelPark Arenberg, 10, B-3001 Leuven, Belgium, and the University of Cambridge, Department of Engineering, Trumpington Street, CB2 1PZ, {\tt\small rodolphe.sepulchre@kuleuven.be}.}
\thanks{The research leading to these results has received funding from the European Research Council under the Advanced ERC Grant Agreement SpikyControl n.101054323.}%
}

\begin{document}

\maketitle
\thispagestyle{empty}
\pagestyle{empty}

\begin{abstract}
This paper proposes a variable metric splitting algorithm to solve the electrical behavior of neuromorphic circuits made of capacitors, memristive elements, and batteries. The gradient property of the memristive elements is exploited to split the current to voltage operator as the sum of the derivative operator, a Riemannian gradient operator, and a nonlinear residual operator that is linearized at each step of the algorithm. The diagonal structure of the three operators makes the variable metric forward-backward splitting algorithm scalable and amenable to the simulation of large-scale neuromorphic circuits.  

\end{abstract}

\section{Introduction}

Neuromorphic circuits are receiving increasing attention for their potential in making computation and signal processing more energy-efficient and event-driven rather than clock-driven \cite{khacef2023spike}. 

A major current bottleneck of this emerging discipline lies in the lack of a {\textit{scalable}} simulation, analysis, and design framework. Any progress in making the theory of neuromorphic circuits more scalable is of also of significance for computational neuroscience, which faces the same obstacles when modeling large-scale neuronal populations.

Both neuromorphic circuits and the biophysical models of computational neuroscience consist of three types of circuit elements: capacitors, batteries, and {\it conductance-based} or, equivalently, {\it memristive} elements \cite{chua1971memristor}. The latter are current sources that obey the Ohmic law of resistive elements, but with a conductance that has memory, that is, depends on the past history of the applied voltage. This voltage-dependence of the conductance was identified in the seminal work of Hodgkin and Huxley \cite{hodgkin1952quantitative} as a key biophysical mechanism enabling the excitability properties of neurons.  For this reason, memristive elements are fundamental to the spiking behavior of neuromorphic circuits and neuronal behaviors \cite{sepulchre2022spiking}. 

In the previous work \cite{shahhosseini2024operator}, we explored the potential of operator-theoretic splitting methods to solve the behavior of neuromorphic circuits. This work builds upon the earlier PhD dissertation \cite{chaffey2022input} that highlighted the value of splitting algorithms to solve circuit equations that interconnect monotone elements. Maximal monotonicity is well-known to lie at the core of large-scale optimization algorithms \cite{ryu2022large}, but the work of \cite{chaffey2023monotone} highlighted the fact that maximal monotonicity was first introduced in circuit theory to characterize nonlinear resistors, and that splitting algorithms are particularly adjusted to exploit the topology of circuits composed of monotone elements. 

The goal of the present paper is to show the value of connecting maximal monotonicity to memristive modeling and how this connection might be a game changer for the simulation of neuromorphic circuits. The link between maximal monotonicity and memristive modeling is through the companion article \cite{Forni2024memristive} that revisits memristive modeling as gradient modeling in the space of past signals: in the same manner that a scalar conductor $i=gv$ can be regarded as the gradient of the quadratic $<v,v>$ in $\mathbb{R}$, with respect to the non-euclidean metric $<x,y>=g^{-1}xy$, the same is shown to hold for a memristive element $i(t)=g(\bold v)(t)v(t)$ where the quadratic $<\bf v,\bf v>$ is now in the signal space of past signals. The maximal monotonicity of a Riemannian gradient operator can also be defined with respect to a variable metric. The consequence of this link between monotonicity and memristive modeling is that, from that generalized perspective,  any neuromorphic circuit becomes an interconnection of monotone elements (with respect to different metrics), which brings neuromorphic circuits much closer to the framework considered in \cite{chaffey2023monotone}. 

This new development is a departure from our earlier work \cite{shahhosseini2024operator}, where each memristive element was replaced by a nonlinear amplifier (in the same way as nonlinear resistors in the FitzHugh-Nagumo model approximate the memristive elements of the Hodgkin-Huxley model). This apparent ``simplification" was a bottleneck of our earlier work \cite{shahhosseini2024operator} because it resulted in circuits whose individual elements were not monotone and had to be modeled as differences of monotone operators. In contrast, with the architecture proposed in the present paper, only the constant batteries are non-monotone, resulting in a highly structured and tractable departure from an otherwise monotone formalism.

The rest of the paper is organized as follows. Section 2 provides an exposition to operator-theoretic approaches, discussing their advantages and limitations. Section 3 reviews the structure and properties of the memristive model of neuromorphic systems, highlighting its minimal formulation. Section 4 discusses variable metric splitting methods and introduces the variable-metric forward-backward splitting algorithm. Section 5 showcases the strength and capabilities of this method through two network examples. Section 6 concludes the paper, addressing the immediate future steps.

\section{Operator-Theoretic Representation of Neuromorphic Systems}

\subsection{Mathematical Preliminaries}

This subsection introduces a few important mathematical definitions that are used throughout this paper. $\mathcal{L}^2_\mathbb{T}$ denotes the space of square-integrable signals over the time axis $\mathbb{T}$, equipped with the inner product
\begin{equation}
    \ip{u}{y} = \int_{\mathbb{T}} u^\top(t)y(t)dt < \infty
\end{equation}

Notation $\mathcal{L}^2$ is used to denote $\mathcal{L}^2_{[0, \infty)}$, and $\mathcal{L}^2_T$ to denote $\mathcal{L}^2_{[0, T]}$. The latter describes the space of $T$-periodic signals, restricted to a single period. The space $\mathcal{L}^2 _T$ is defined for continuous signals, and the space of discrete-time square-integrable signals is denoted as $ l^2 _T$.

\textbf{Definition 1.} Operator $\operatorname{A}$, defined on the space $\mathcal{L}^2$, describes a mapping $\operatorname{A}: \mathcal{L}^2 \rightarrow \mathcal{L}^2$ and its {\textit{graph}} is defined as
\begin{equation}
    \text{Gra} \operatorname{A} = \{(u,y) | y \in \operatorname{A}(u)\} \subseteq \mathcal{L}^2 \times \mathcal{L}^2.
\end{equation}

\textbf{Definition 2.} Operator $\operatorname{A}: \mathcal{L}^2 \rightarrow \mathcal{L}^2
$ is {\textit{monotone}} if 
\begin{equation}
    \ip{u_1 - u_2}{y_1 - y_2}  \geq 0
\end{equation}
for all $u_1, u_2 \in \dom {\operatorname{A}}$ and $y_1, y_2$ are the corresponding outputs. Operator $\operatorname{A}$ is defined to be \emph{maximal monotone} if no other monotone operator’s graph contains its graph.

\textbf{Definition 3.} The \emph{resolvent} operator of $\operatorname{A}: \mathcal{L}^2 \rightarrow \mathcal{L}^2$, denoted by $\operatorname{J}_{\alpha \operatorname{A}}: \mathcal{L}^2 \rightarrow \mathcal{L}^2$ is defined as
\begin{equation}
    \operatorname{J}_{\alpha \operatorname{A}} = (I + \alpha \operatorname{A})^{-1}
\end{equation}
where $\alpha$ is a scalar and $\alpha > 0$.

\subsection{Exposition of the Operator-Theoretic Approach}{\label{subsec_OTA}}

In recent years, there has been a resurgence of interest in operator-theoretic methods for simulating and analyzing nonlinear dynamical systems \cite{chaffey2023monotone}. This subsection delineates the motivation for this approach, focusing on why it is an adequate choice for the class of neuromorphic systems, and continues with an overview of the methodology.

\subsubsection{Motivation}

The operator-theoretic framework not only addresses several limitations that numerical integration methods face in solving neuromorphic systems but also provides a more efficient and flexible approach that cannot be attained otherwise. The following are among the most prominent motivations.

\begin{enumerate}[I.]
    \item Efficient Variability Analysis: In an operator-theoretic representation, the input ($i$) and output ($v$) are related through
    \begin{equation}{\label{eq:OTR}}
        \operatorname{G}(v) = i 
    \end{equation}
    where the continuous operator $\operatorname{G} : \mathcal{L}^2 \rightarrow \mathcal{L}^2$ characterizes the dynamical system of interest. This formulation not only defines a continuous mapping between the input and the output but also provides a continuous dependence of the output on system parameters, reflected through the structure of $\operatorname{G}$. Consequently, infinitesimal changes in the parameters lead to an output signal that remains close to the original output. This continuity enables an efficient study of behavioral variability by systematically sweeping through broad parameter ranges.

    \item Event-Oriented Functionality: The multi-timescale essence of excitable behavior is realized by {\textit{stiff}} nonlinear ODEs. This stiffness forces numerical integration methods to use small stepsizes, resulting in heavy computational loads and, for large networks, intractability. Additionally, poor stepsize selection can lead to solver instability or erroneous evaluations that propagate through the solution, undermining the simulation’s validity. In sharp contrast, operator-theoretic methods do not compute the solution incrementally but instead derive the entire solution signal in one step. As a result, they are less affected by stiffness. Furthermore, even with a low resolution, operator-theoretic methods can identify the key events that define the system's behavior, though they may not capture every detail with full accuracy. This capability makes operator-theoretic methods particularly well-suited for handling the event-based dynamics inherent in neuromorphic systems.

    \item Scalability and Adjustability: The operator-theoretic methods used in simulating spiking systems \cite{shahhosseini2024operator} rely on first-order methods of convex optimization \cite{ryu2022large}. These methods are known for their computational tractability and scalability. Additionally, the operator-theoretic framework offers the flexibility to choose the signal space \cite{chaffey2023monotone}. While $\mathcal{L}^2$ has been the traditional choice, the flexibility to choose different signal spaces allows the user to adjust the level of detail in the simulation, optimizing the trade-off between computational efficiency and the precision of the modeled behavior.

\end{enumerate}

\subsubsection{From Concept to Methodology}

Any neuromorphic system can be realized through the operator-theoretic representation of (\ref{eq:OTR}). Given a fixed input $i^* \in \mathcal{L}^2$, it is possible to rewrite the dynamics as
    \begin{equation}{\label{eq:ZFP}}
        \operatorname{G}(v) - i^* = 0 
    \end{equation}
and thus, the task of simulating and analyzing neuromorphic systems boils down to the problem of finding the solution signal $v^*$ that corresponds to the zero of operator formulation of (\ref{eq:ZFP}).

The problem of finding the zero of {\textit{monotone}} operators is well-established in the literature. These methods are particularly prominent due to the central role of monotonicity in first-order methods of convex optimization, as the subgradient of any convex function is monotone. These methods rely on transforming the monotone zero finding problem (ZFP) into a contractive (or averaged) fixed-point iteration (FPI) algorithm. Additionally, monotonicity serves as a bridge between algorithmic tractability and physical interpretability, with the monotonicity of an element corresponding to its incremental passivity.

The richness of spiking dynamics gives rise to a complicated operator $\operatorname{G}(\cdot)$, making it computationally infeasible to directly solve the problem outlined in (\ref{eq:ZFP}). To address this challenge, operator theory introduces splitting algorithms that decompose the main operator into simpler sub-operators, enabling iterative steps based only on these sub-operators. One prominent example is the Douglas-Rachford Splitting (DRS) algorithm, which corresponds to the Alternating Direction Method of Multipliers (ADMM) \cite{boyd2011distributed}. However, splitting a complex operator like $\operatorname{G}(\cdot)$ into sub-operators is not always straightforward and does not necessarily yield a unique solution. The complexity of the original operator, combined with the non-uniqueness of splitting, complicates the task of finding an effective splitting for neuromorphic systems, making it a challenge in this context.

The essence of excitable behavior stems from the interplay of monotone and anti-monotone elements at different timescales \cite{ribar2019neuromodulation}. Thus, the operator-theoretic representation of neuromorphic systems can never be realized by purely monotone elements. Nevertheless, it is possible to reformulate the problem to a difference-of-monotone (DM) structure. This permits the use of ideas and methods from difference-of-convex algorithms (DCA) \cite{shen2016disciplined} and concave-convex programming (CCP) \cite{yuille2001concave}. As a direct consequence, DM splitting algorithms have to be employed for the objective of simulating neuromorphic systems.

\subsection{Neurophysiology-Compliant Neuromorphic Models}

Among the plethora of models used to emulate the behavior 
 of neurons, conductance-based models \cite{izhikevich2007dynamical} hold a special place. They conform to neurophysiology and are true to the biophysics of the problem. In contrast, mathematical models (e.g., leaky-integrate and fire), that mostly come from the simplification and reduction of conductance-based models, are known to lead to erroneous predictions \cite{burger2023oscillatory}. Thus, it is important to understand the underlying mechanisms generating the excitable behavior of conductance-based models and preserve it.

\subsubsection{Conductance-based Models and Excitability}

The seminal work of Hodgkin and Huxley \cite{hodgkin1952quantitative} explained the biological mechanism of excitability through the interaction of ion channels with voltage-gated conductances. Later, reduced-order models \cite{fitzhugh1961impulses} distilled the essence of this mechanism. These models demonstrated that excitability arises when elements with negative incremental conductance (such as $\text{Na}^+$ in the Hodgkin-Huxley model) dominate at fast timescales, while elements with positive incremental conductance (like $\text{K}^+$) control the dynamics at slower timescales. Recently, this mechanism has been revisited through the lens of the mixed-feedback structure \cite{sepulchre2022spiking}, which bridges neuromorphic models and system theory.

\subsubsection{Building Blocks of Neuromorphic Behavior}

To create excitable behaviors from scratch, three archetypes of blocks are needed. First is a passive block that represents the capacitor in the circuit model. This block attempts to emulate the effect of the membrane in biological neurons. The second block is the element with negative incremental conductance (at specific voltage ranges) that provides the system with (localized) positive feedback. This block is the destabilizing agent and creates the upward stroke of the spike. The last block represents the positive incremental conductance elements that correspond to negative feedback. These blocks regulate behavior and stabilize the system after a spike has occurred. The interested reader is referred to \cite{ribar2019neuromodulation}, which discusses this mechanism using these three types of blocks in detail.

\subsection{The Bottleneck of Neuromorphic Modeling}

Encapsulating excitable behaviors using the three aforementioned archetype blocks is consistent with the essence of excitability but comes with certain limitations. First, the elements of positive and negative incremental conductance do not necessarily demonstrate monotonic behavior. This necessitates the decomposition of these elements into the difference of two monotone operators so that splitting algorithms can be used. This is neither a computationally simple task nor does it have strong neurophysiological grounds. Additionally, and more importantly, the synaptic connections in this setup are never monotone. For a network of $n$ neurons, this requires the decomposition of the operator of the synaptic connection into at least $2n$ operators to preserve its compatibility with the structure of consensus-based splitting methods. Nevertheless, this leads to a linear enlargement of the consensus set and significantly decelerates convergence. The slowdown of the convergence rate creates the scalability bottleneck. This motivates an alternative formulation of conductance-based models that not only maintains its conformity with neurophysiology but is also superb for use in {\textit{scalable}} operator-theoretic solvers.

\section{Memristive Modeling of Neuromorphic Systems}

\subsection{Structure of the Model}

We consider neuromorphic circuits that closely mimic the architecture of the conductance-based models of neurophysiology. Such models have a very particular structure. Each neuron is a memRC circuit that interconnects a capacitor with a parallel bank of internal and external current sources. The internal current sources model the ion channels that gate the flow of currents in the cell. The external current sources model the flow of currents resulting from synaptic interconnections with other neurons. Both internal and external current sources obey Ohmic law
$$i=g (v-v_{0})$$ 
which can be regarded as the series interconnection of a constant voltage battery $v_0$ with a memristive current $i=gv$. Memristive means that the conductance $g$ at time $t$ depends on the past history of the voltage $v$. For internal currents, the voltage dependence of $g$ is only on the membrane voltage $v$. For external currents, the voltage dependence of $g$ is on the membrane voltage of other neurons in the circuit.

We first consider the model of a single neuron with membrane voltage $v$. The circuit, in that case, only includes internal currents. All external currents are lumped in the external input $i$.

\begin{equation}{\label{eq:model}}
 c\dot{v} = - \sum _{j=1} ^P g_j (v- N_j) + i_{\text{ext}}(t)  
\end{equation}
where $c$ is the capacitance, $i_{\text{ext}}(t)$ is the externally injected current (the input), and $P$ is the number of memristive elements. $N_j$ is the Nenrst potential corresponding to the $j^\text{th}$ branch and determines the excitatory or inhibitory role of the branch in the internal neural dynamics.

For the sake of illustration in this paper, we will model the voltage dependence of the conductance $g_j$ via a simple convolution operator: 
\begin{equation}{\label{eq:model2}}
\begin{aligned}
    &
  g_j(v_x) = \bar{g}\; \text{ReLU}(v_x-v^{th}) \\
                & \tau _x \dot{v}_x = v - v_x 
                \end{aligned}
\end{equation}
where $x$ denotes the timescale of the filtered voltage, namely instantaneous, fast, slow, and ultraslow, where this is dictated by $\tau _x$.  The nonlinear readout uses a Rectified Linear Unit ($\text{ReLU}$), and $v^\text{th}$ denotes the threshold of the conductance, dictating the voltage range of its activity. $\bar{g}$ denotes the maximal conductance. More complicated models of conductance can be considered, but this complexity is irrelevant for the present paper. What matters is the evaluation of the  conductance as an operator of the voltage past history.

The model of (\ref{eq:model}) can be reformulated to the more convenient format of
\begin{equation}{\label{eq:EI_Decomposition}}
        c\dot{v} = - \sum _{j=1} ^{P_E} g_j(v_x) (v- E_j)  - \sum _{j=1} ^{P_I} g_j(v_x) (v- I_j) + i_{\text{ext}}(t)
 \end{equation}
where the first ${P_E}$ series of memristors are coupled with the positive Nerst potentials $E_j$ and excite the dynamics and the next ${P_I}$ series of memristors are coupled with the negative Nerst potentials $I_j$ and inhibit the dynamics. Bear in mind that $P_I + P_E = P$.

 As explained in the introduction and further detailed in the companion paper \cite{Forni2024memristive}, the memristive operator $i=gv$ can be regarded as the gradient of a quadratic functional $<v,v>$ with respect to a Riemannian metric determined by the inverse of the memductance. In fact, the model of (\ref{eq:model}) can be reformulated to
\begin{equation}{\label{eq:grad}}
    c\dot{v} = - \frac{1}{2} \sum _{j=1} ^P \text{grad}(\ip{v}{v-N_j}) + i_{\text{ext}}(t)
\end{equation}
where the gradient operator of an arbitrary functional $f(x)$, with $x \in \mathcal{L}^2$ and the space equipped with a variable metric $M(x)$, is defined as
\begin{equation}{\label{eq:gradient}}
    \text{grad}(f(x)) = M^{-1}(x) \nabla f(x)
\end{equation}
and by contrasting (\ref{eq:model}), (\ref{eq:grad}) and (\ref{eq:gradient}), it is evident that the variable metrics that define the gradient of the quadratic functions of (\ref{eq:grad}) are the inverse of the conductances of the memristive elements of (\ref{eq:model}), or mathematically
\begin{equation}{\label{eq:equivalency}}
    g_j(v) = M^{-1}_j(v)
\end{equation}

where in addition to the physical elegance of this representation, its structure provides computational advantages that pave the way for the scalable simulation of neuromorphic systems.

The operator-theoric representation of (\ref{eq:model}) corresponds to 
\begin{equation}{\label{eq:model_OT}}
\begin{aligned}
    & c\operatorname{D}v = - \sum _{j=1} ^P \operatorname{G}_j(v_x) (v- N_j) + i_{\text{ext}}\\
    & v_x = (\tau _x \operatorname{D} + \operatorname{Id})^{-1}v
    \end{aligned}
\end{equation}
where $i_{\text{ext}} \in \mathcal{L}^2$, the operator $\operatorname{D} : \mathcal{L}^2 \rightarrow \mathcal{L}^2$ defines the differentiation operator, $\operatorname{Id} : \mathcal{L}^2 \rightarrow \mathcal{L}^2$ defines the identity operator and $\operatorname{G}_j(\cdot) : \mathcal{L}^2 \rightarrow \mathcal{L}^2$ is the operator representation of the conductance $g_j$. Given (\ref{eq:equivalency}), the operator-theoretic representation of (\ref{eq:model_OT}) is equivalent to the operator ZFP of
\begin{equation}{\label{eq:model_metric}}
    c\operatorname{D}v + \sum _{j=1} ^P M_j^{-1}(v_x) (v- N_j) - i_{\text{ext}} = 0
\end{equation}
and thus, the model can be seen as the summation of a differentiator with monotone affine operators defined with respect to variable metrics. 

A {\textit{critical}} point here is that the complexity of neuromorphic dynamics is absorbed into the conductance (or equivalently, the variable metric), while the operators themselves remain extremely simple. This absorption of complexity into the conductance is the key property that enables minimal splitting, ultimately allowing for scalability.

\textit{Remark}: There is an important technical difference between $v$ in the state-space representation of (\ref{eq:model}) and operator-theoretic representation of (\ref{eq:model_OT}). In the representation of (\ref{eq:model}), $v$ is a state variable whereas in (\ref{eq:model_OT}), $v$ is a signal in $\mathcal{L}^2$.

\subsection{Reformulation of the Memristive Model}

The ZFP of (\ref{eq:model_OT}) must be split into sub-operators to become a computationally tractable problem. This can be achieved by rewriting (\ref{eq:model_OT}) as
\begin{equation}{\label{eq:internal}}
    c\operatorname{D}v + \left(\sum _{j=1} ^P \operatorname{G}_j(v_x)\right) v - \left( (\sum _{j=1} ^P \operatorname{G}_j(v_x) N_j) + i_{\text{ext}} \right) = 0
\end{equation}
where it is possible to define the new parameter $\operatorname{G}_{\text{tot}}^{{\text{int}}}(v) = \sum _{j=1} ^P \operatorname{G}_j(v_x)$ and also $N_{\text{tot}}^{\text{int}}(v) = \sum _{j=1} ^P \operatorname{G}_j(v_x) N_j$ the representation further simplifies to 
\begin{equation}{\label{eq:simplified}}
    c\operatorname{D}v + \operatorname{G}_{\text{tot}}^{{\text{int}}}(v) v - \left( N_{\text{tot}}^{\text{int}}(v) +i_{\text{ext}} \right) = 0
\end{equation}
where this can now be seen as a three-operator splitting problem with variable metrics and offsets. The first two operators of (\ref{eq:simplified}) are monotone (though with a variable metric), but the third operator $N_{\text{tot}}^{\text{int}}(v) +i_{ext}$ entails both monotone and anti-monotone elements. Using the concept of the concave-convex procedure \cite{yuille2001concave}, it is possible to simplify this operator and use it within the structure of DM algorithms. The $\text{int}$ superscript indicates that these are the {\textit{internal}} conductances, coming from the internal ionic currents of the neuron and not the synaptic connections.

\subsection{Synaptic Connections}

The synaptic connections follow a similar structure to the ionic currents of the individual neurons. An excitatory synaptic connection from a pre-synaptic neuron $N_{\text{pre}}$ to a post-synaptic neuron $N_{\text{post}}$ is characterized as 
\begin{equation}{\label{eq:syn}}
    i_{\text{syn}} =  \underset{G_{\text{syn}}(v^{\text{pre}})}{\underbrace{\bar{g}_{\text{syn}}^{N_{\text{pre}} \rightarrow N_{\text{post}}} \text{ReLU}(v_x^{\text{pre}}-v^{th}_E)}} (v^{\text{post}}-E_{\text{syn}})
\end{equation}
where $\bar{g}_{\text{syn}}^{N_{\text{pre}} \rightarrow N_{\text{post}}}$ is the maximal conductance and ${N_{\text{pre}} \rightarrow N_{\text{post}}}$ indicates that the current flows from the pre-synaptic neuron into the post-synaptic. This means that the pre-synaptic neuron's voltage gates the conductance of the synaptic connection, and upon having a spike, the conductance permits the flow of currents that excite the post-synaptic neuron. Similarly, for inhibitory connections, it is possible to write
\begin{equation}{\label{eq:syn_I}}
    i_{\text{syn}} =  \underset{G_{\text{syn}}(v^{\text{pre}})}{\underbrace{\bar{g}_{\text{syn}}^{N_{\text{pre}} \rightarrow N_{\text{post}}} \text{ReLU}(v_x^{\text{pre}}-v^{th}_I)}} (v^{\text{post}}-I_{\text{syn}})
\end{equation}
where the only change is with the Nerst potential, from $E_{\text{syn}}$ to $I_{\text{syn}}$ and the threshold voltage $v^{th}_I$.

It is noteworthy that the structure of the synaptic currents is identical to the structure of internal ionic currents, and thus, it is straightforward to absorb synaptic currents in the structure of (\ref{eq:simplified}). To provide an elucidating example, the dynamics of a neuron with $Q$ excitatory pre-synaptic neurons can be written using (\ref{eq:internal}) and (\ref{eq:syn}) as
\begin{equation}
\begin{aligned}
    & c\operatorname{D}v + \left(\sum _{j=1} ^P \operatorname{G}_j(v_x)\right) v - \left( (\sum _{j=1} ^P \operatorname{G}_j(v_x) N_j) + i_{\text{ext}} \right) \\
    & + \sum _{k = 1} ^Q G_{\text{syn}}(v^{\text{pre},k}) (v-E_{\text{syn}}) = 0
    \end{aligned} 
\end{equation}
where the last sum accounts for the effect of the synaptic connections and the $v^{\text{pre},k}$ identifies the voltage of the $k^{\text{th}}$ pre-synaptic neuron. This expression can be reformulated to

\begin{equation}
\begin{aligned}
    & c\operatorname{D}v + \left(\sum _{j=1} ^P \operatorname{G}_j(v_x)  + \sum _{k = 1} ^Q G_{\text{syn}}(v^{\text{pre},k})\right) v \\
    & - \left( (\sum _{j=1} ^P \operatorname{G}_j(v_x) N_j) + ( \sum _{k = 1} ^Q G_{\text{syn}}(v^{\text{pre},k})E_{\text{syn}}) + i_{\text{ext}} \right) = 0
    \end{aligned} 
\end{equation}
where by setting $G_{\text{tot}} = \sum _{j=1} ^P \operatorname{G}_j(v_x)  + \sum _{k = 1} ^Q G_{\text{syn}}(v^{\text{pre},k})$ and $N_{\text{tot}} = (\sum _{j=1} ^P \operatorname{G}_j(v_x) N_j) + ( \sum _{k = 1} ^Q G_{\text{syn}}(v^{\text{pre},k})E_{\text{syn}})$ we recover the exact form of (\ref{eq:simplified}), this time for the entire network as
\begin{equation}{\label{eq:ntework}}
    c\operatorname{D}v + \operatorname{G}_{\text{tot}}(v) v - \left( N_{\text{tot}}(v) +i_{\text{ext}} \right) = 0
\end{equation}
where now the neuron and its entire synaptic network is defined through three operators. Solving this three-operator ZFP provides the solution of this network.

\section{Methodology}

The operator-theoretic representation of neuromorphic systems, denoted in (\ref{eq:ntework}), can be regarded as a variable metric splitting problem with two monotone operators and a third operator that is approximated and treated as an offset. This idea is widespread in DCA and DM splitting algorithms, such as the Difference-of-Monotone Douglas-Rachford algorithm \cite{chuang2022unified}. This section discusses variable metric methods and utilizes the variable metric forward-backward splitting (VMFBS) to solve the ZFP of (\ref{eq:ntework}).

\subsection{Variable Metric Forward-Backward Splitting Algorithm}

To solve the operator ZFP of the form

\begin{equation}
    (\operatorname{A} + \operatorname{B} - \operatorname{C})v=0
\end{equation}
where $v \in \mathcal{L}^2$, operator $\operatorname{A} : \mathcal{L}^2 \rightarrow \mathcal{L}^2$ is monotone, $\operatorname{B} : \mathcal{L}^2 \rightarrow \mathcal{L}^2$ is monotone with a metric defined as $M(v)$ and $\operatorname{C}$ is an operator that will be approximated as a constant offset in each iteration, it is possible to use the VMFBS algorithm. The pseudo-code for the algorithm can be seen as

\begin{algorithm}
	\caption{VMFBS} 
	\begin{algorithmic}[1]
		\For {$k=1,2,\ldots, \text{max-iteration}$} 
                \State $z^{k+1}= v^k - \alpha M^{-1}(v^k) \operatorname{B}(v^k) + \alpha \operatorname{C}(v^k)$
                \State $v^{k+1}=\operatorname{J}_{\alpha \operatorname{A}}(z^{k+1})$
		\EndFor
	\end{algorithmic} 
\end{algorithm}
where $\operatorname{J}_{\alpha \operatorname{A}}(\cdot)$ is the resolvent operator of operator $\operatorname{A}$.

In this representation, operator $\operatorname{A}(\cdot)$, $\operatorname{B}(\cdot)$ and $\operatorname{C}(\cdot)$ correspond to $c \operatorname{D}(\cdot)$, $\operatorname{Id}(\cdot)$ and $N_{\text{tot}}(\cdot) +i_{\text{ext}}$ of (\ref{eq:ntework}). In this case, the inverse of the variable metric of operator $\operatorname{B}(\cdot)$, that is $M$, is the total conductance $\operatorname{G}_{\text{tot}}$ of the memristive network.

An interesting remark is the tight connection of the VMFBS algorithm with Riemannian Proximal Gradient methods \cite{huang2022riemannian}. This provides a conceptual link to the re-interpretation of the conductances in the memristive modeling language of this paper. Here, with this new interpretation, the conductances can be thought of as Riemannian metrics that affect the space within which the quadratic functionals are defined. This also permits the treatment of neuromorphic systems as manifold optimization problems.

\subsection{Adapting Splitting Methods to Neuromorphic Systems}{\label{subsec_EDN}}

Using the VMFBS algorithm to simulate neuromorphic systems requires careful attention to key technical aspects. The simulation always starts and ends at the system's resting potential, ensuring that simulated events result solely from external excitation, aligning with neurophysiological principles. 

A crucial computational detail is evaluating the resolvent of the differentiation operator \( \operatorname{D} \). This can be done in two ways. One approach forms the inverse of the resolvent \( (\operatorname{Id} + \alpha \operatorname{D}) \) and transforms it from the time to the frequency domain. Exploiting the linear time-invariant (LTI) structure allows diagonalization, avoiding costly inversion and reducing complexity to \( \mathcal{O}(n \log n) \). This time-frequency hopping is discussed in detail in \cite{shahhosseini2024operator}. This method also fine-tunes spiking behavior by selecting the FFT frequency bandwidth. Alternatively, the resolvent can be evaluated in the time domain using efficient methods like the Thomas algorithm, which has a lower complexity of \( \mathcal{O}(n) \) but does not allow straightforward bandwidth truncation.

The choice of the VMFBS algorithm over other variable-metric splitting methods (e.g., variable-metric Douglas-Rachford) is driven by computational feasibility. Running a backward splitting step on an operator with a variable metric introduces significant complexity due to the presence of voltages at different timescales in total conductance. This leads to implicit, nonlinear, and dynamic problems, making direct computation impractical.

\section{Simulation and Discussion}

The simulation section starts with a simple two-neuron network as a tutorial on using this method. Next, it presents a Pyramidal-Interneuronal Network Gamma (PING) model \cite{borgers2005background}, which plays a key role in rhythmic biological behaviors, highlighting the framework’s strengths and capabilities.

\subsection{Simulation of the E-I motif}

The neuromorphic system of interest here consists of two neurons. The first neuron, denoted by $E$, has an excitatory synaptic connection to the second neuron. The second neuron, denoted by $I$, is connected to the first neuron with an inhibitory connection. The dynamics can be written as
\begin{equation}
\begin{aligned}
    &c \operatorname{D}v^E + g_0v^E + g_1(v^E)(v^E-E) + g_2(v_s^E)(v^E - I) + \\ 
    & g_{\text{syn}}(v_s^I)(v^E-I_{\text{syn}}) - i_{\text{ext}} = 0, \\
    &c \operatorname{D}v^I + g_0v^I + g_1(v^I)(v^I-E) + g_2(v_s^I)(v^I - I) + \\ 
    & g_{\text{syn}}(v_E)(v^I-E_{\text{syn}}) - i_{\text{ext}} = 0 \\
    & v_s^I = (\tau _s \operatorname{D} + \operatorname{Id})^{-1}v^I \\
    & v_s^E = (\tau _s \operatorname{D} + \operatorname{Id})^{-1}v^E \\
\end{aligned}
\end{equation}
with $c=1$, $\tau _s = 10$, $g_1(v) = \text{ReLU}(v-1)$, $g_2(v) = 10 \text{ReLU}(v-1)$, $g_{\text{syn}}(v) = 1.5\text{ReLU}(v-1)$, $E=10$, $I=-10$, $E_{\text{syn}}=10$ and $I_{\text{syn}}=-10$ and the $E$ neuron is excited with a square-wave of amplitude $0.15$ that starts at $t_{\text{start}}=2 ms$ and ends at $t_{\text{start}}=30 ms$.

\begin{figure}
    \centering
    \includegraphics[width=1\linewidth]{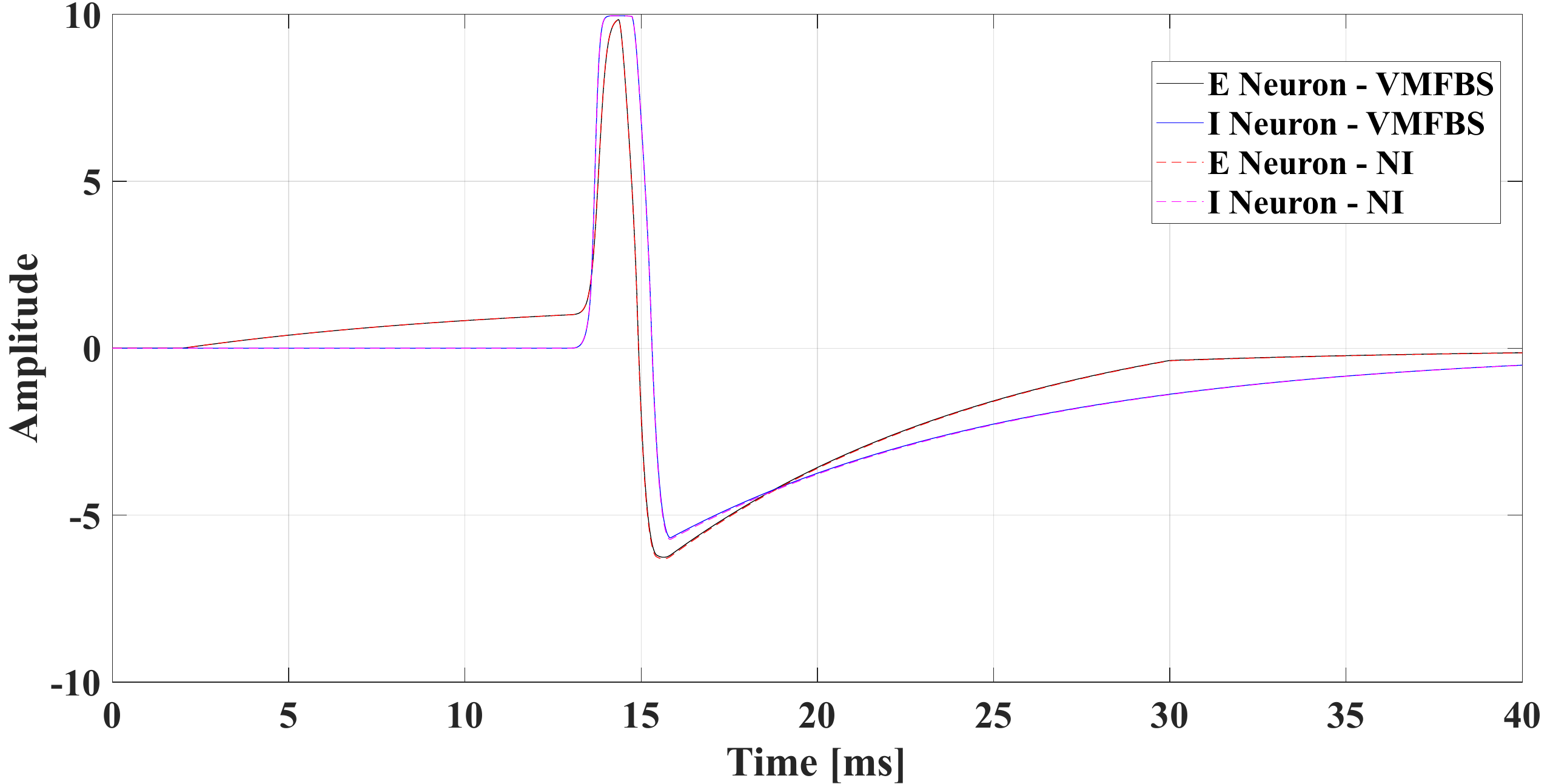}
    \caption{The simulation results of the two-neuron E-I motif using VMFBS algorithm and Adams-Bashforth two-step method.}
    \label{fig:EI_Simple}
\end{figure}

In this example, the neurons are identical, and the only difference is in the synaptic connection. In this scenario, both neurons start from rest, but the $E$ neuron receives an externally injected current. As a result, this neuron spikes. The spiking of the $E$ neuron excites the $I$ neuron through the excitatory synapse, and thus, the $I$ neuron also fires. Simulating this network with the operator-theoretic framework of this paper results in the illustrations of Fig \ref{fig:EI_Simple} where the results match with that of NI methods.

\subsection{Simulation of the PING Network}

We now move from one motif of two neurons to a so-called E-I network, which serves as the backbone for many architectures in neuroscience, including the PING rhythm \cite{borgers2005background}. 

\begin{figure}[b]
    \centering
    \includegraphics[width=1\linewidth]{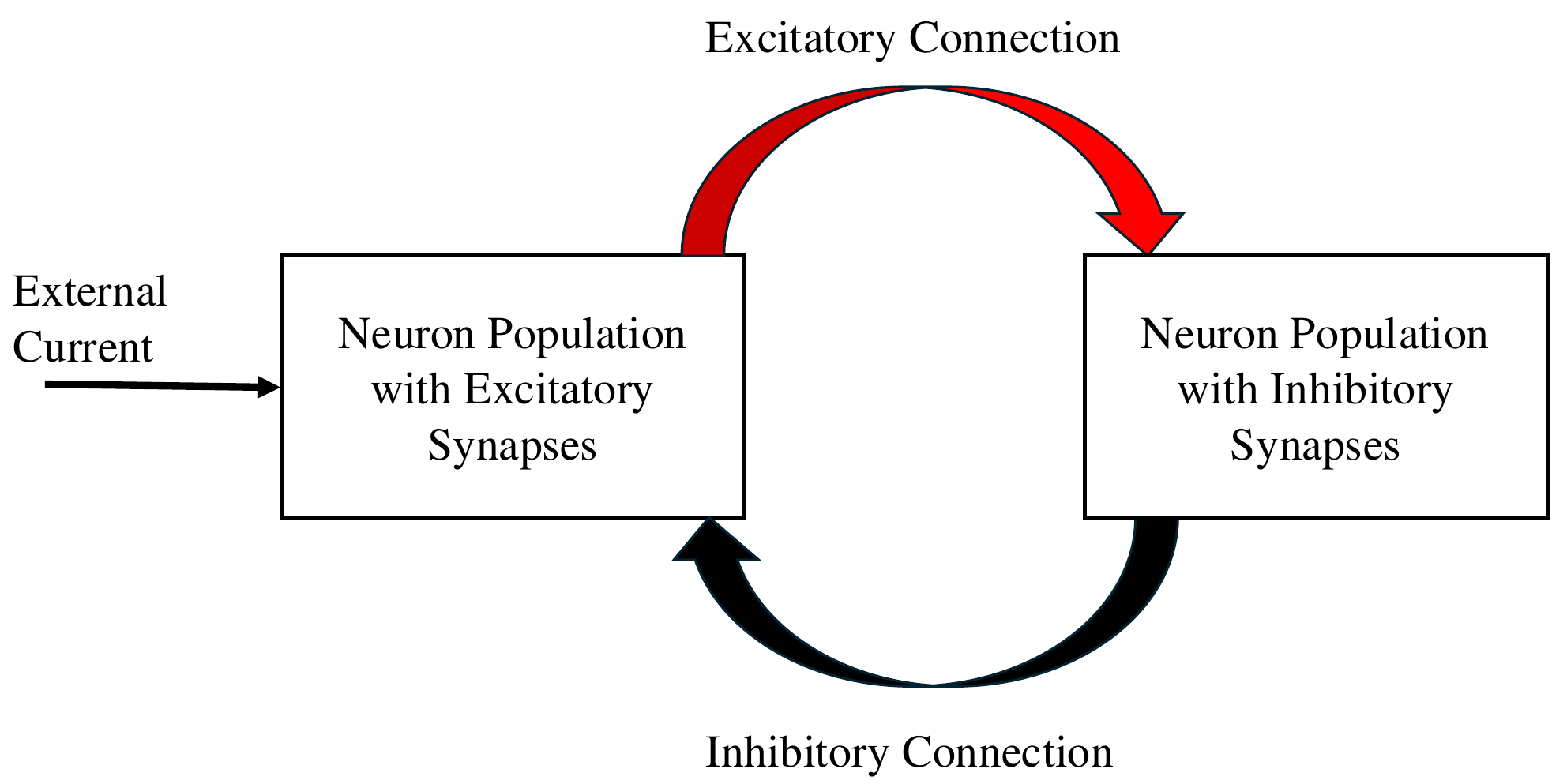}
    \caption{The network consists of two populations of neurons, with no synaptic connections within each population. The external current excites the first population, and the excitatory synapses propagate this excitation to the second population. The excitation and spiking of the second population inhibits the first population, creating the well-known E-I motif rhythms.}
    \label{fig:EI_Network}
\end{figure}

This example attempts to demonstrate a network-level behavior. Here, we examine a network of 50 neurons, where the parameters corresponding to the internal dynamics of the individual neurons are as follows. $E=20$, $I=-20$, $\tau _s = 5$, and the dynamics of the internal contactances are identical to the previous example. This network is divided into two populations: the first 40 neurons form an excitatory group that projects synaptic connections to the remaining 10 neurons, which serve as inhibitory counterparts. In return, these inhibitory neurons suppress the activity of the excitatory population, creating the classic E-I motif that underlies rhythmic behavior. Figure \ref{fig:EI_Network} demonstrates this network schematically. The connection from the excitatory population to the inhibitory is only in effect after $t_{\text{effect}}=120 ms$, and before that, the excitatory synaptic connections are off. This is added to demonstrate the kick-start of the rhythmic behavior.

\begin{figure}
    \centering
    \includegraphics[width=1\linewidth]{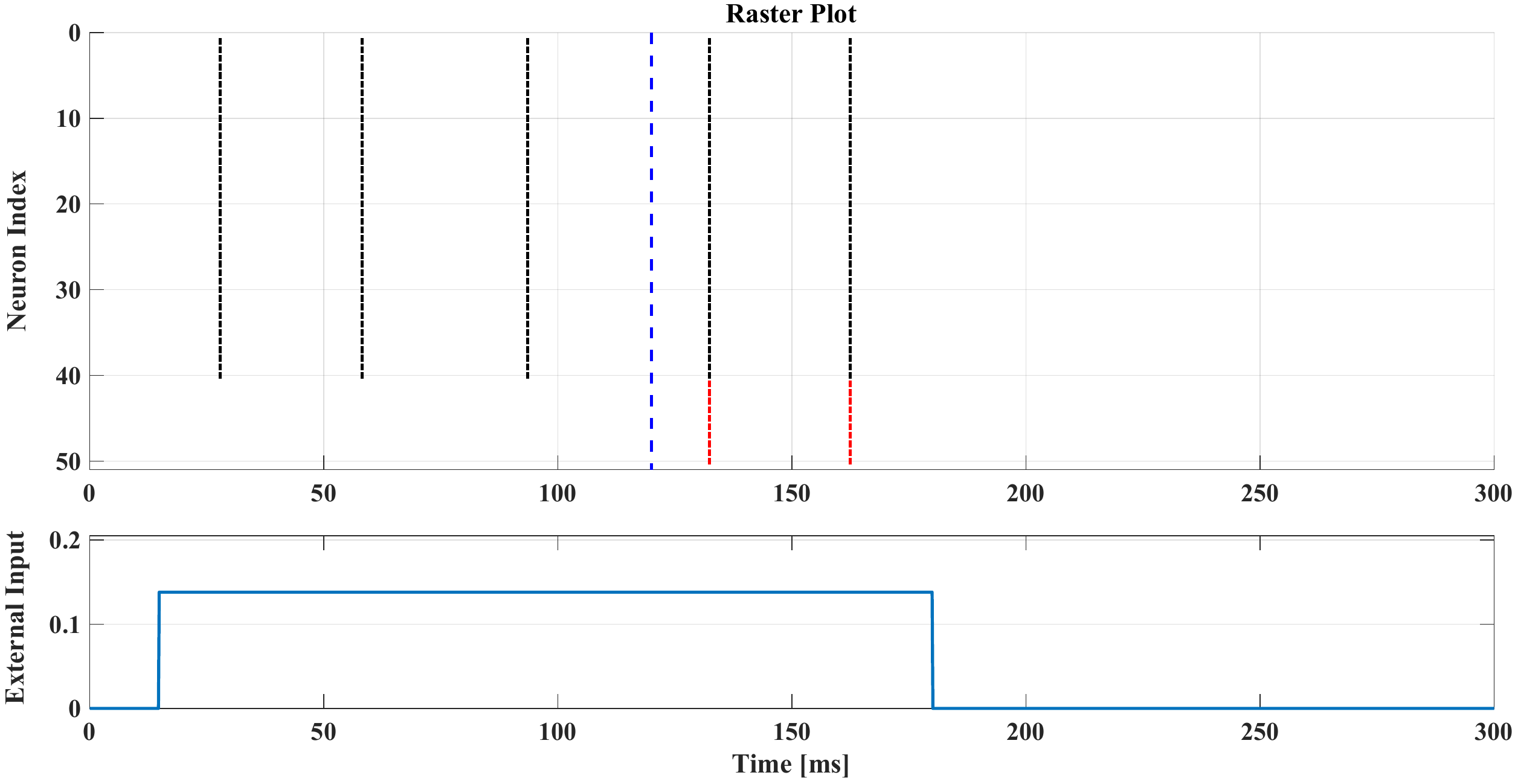}
    \caption{The raster plot of the two population network is presented here. The black dashes indicate the spike of the excitatory population, and the red dashes indicate the spikes of the inhibitory neurons. The excitatory synaptic connection only comes into play after $t_{\text{effect}}=120 ms$, indicated by the blue dashed line. A strong rhythmic behavior can be observed with the given external excitation.}
    \label{fig:final}
\end{figure}

The network is initially at rest, consistent with the event-based formalism on which this simulation framework is built \ref{subsec_EDN}. A square-wave current, as depicted in Fig \ref{fig:final}, is injected into each of the 40 excitatory neurons. This input is expected to activate the excitatory population, triggering spike trains that, in turn, stimulate the inhibitory neurons after $t_{\text{effect}}=120 ms$. As a result, a strongly rhythmic pattern of activity should emerge. Figure \ref{fig:final} demonstrates the result of the simulation of this network with the proposed method using a raster plot \cite{rieke1999spikes}.

The parametric values of the operator-theoretic solver (VBFBS in this case) are chosen to provide reliable and accurate simulations. The externally injected current begins at $t_{\text{start}}=15 ms$ and ends at $t_{\text{end}}=180ms$ with an amplitude of $0.138$. The strength of the maximal excitatory conductance $\bar{g}_{syn}^{E\rightarrow I} = \frac{0.2}{N_E}$ where $N_E=40$ is the number of neurons in the excitatory population. Similarly, $\bar{g}_{syn}^{I\rightarrow E} = \frac{0.4}{N_I}$ where $N_I=10$. The sampling frequency has chosen to be $F_s = 6 Hz$, and the stepsize of the VBFBS is $\alpha = 0.28$. The Nerst potential of the synaptic connections are $E_{syn} = 20$ and $I_{syn} = -20$. The threshold of the excitatory synaptic connections is $v_E ^{th} = 5$, where this value is $v_I ^{th} = 1.2$ for inhibitory connections. The results of this simulation matches that of NI methods, verifying its validity.

\section{Conclusion}

This paper introduces a new operator-theoretic splitting algorithm to solve the electrical behavior of any neuromorphic circuit defined by parallel interconnections of capacitors with memristive elements in series with constant voltage batteries. 

The key novelty of the proposed algorithm is to connect the monotonicity of the memristive elements to their gradient property in the space of past voltages. 

The proposed splitting is between three operators defined as follows: a derivative operator, which is diagonal in the frequency domain; a gradient operator, with respect to a diagonal metric determined by the total memductance of the circuit, and a non-monotone operator that is treated as a constant offset at every iteration.

The resulting algorithm is highly structured and scalable. Future work will further investigate the use of Riemannian gradient algorithms with a rigorous convergence analysis, as well as the possibility of simulating a large-scale network {\it at scale} by modulating the resolution of the memristive modeling in the temporal and spatial domains. Those two important additions will be included in the journal version of this manuscript.


\bibliographystyle{IEEEtran}
\bibliography{ref}

\end{document}